\documentclass[12pt]{article}

\usepackage{CJKutf8}
\usepackage{natbib}

\usepackage[dvips]{graphicx}
\usepackage[margin=1in,top=0.8in]{geometry}
\usepackage{hhline,amssymb,epsfig,amsmath,array,amsthm,color,setspace,titlesec, hyperref, lipsum,enumitem}
\usepackage[toc]{appendix}
\usepackage[T1]{fontenc}
\usepackage[utf8]{inputenc}
\usepackage{authblk}
\usepackage{changes}
\usepackage{mathrsfs}
\usepackage{subfigure}
\usepackage{color, colortbl}
\usepackage{mathtools}
\usepackage{multirow}

\usepackage{caption}
\usepackage{bm}
\usepackage{multicol}

\def\boxit#1{\vbox{\hrule\hbox{\vrule\kern6pt
          \vbox{\kern6pt#1\kern6pt}\kern6pt\vrule}\hrule}}

\def\independenT#1#2{\mathrel{\setbox0\hbox{$#1#2$}%
    \copy0\kern-\wd0\mkern4mu\box0}}
    
\newcommand{\be}{\begin{eqnarray}}
\newcommand{\ee}{\end{eqnarray}}
\newcommand{\ba}{\begin{eqnarray*}}
\newcommand{\ea}{\end{eqnarray*}}

\newtheorem{lemma}{Lemma}
\newtheorem{theorem}{Theorem}

\newtheorem{assumption}{Assumption}

\newcommand{\argmin}{\arg\!\min}

\begin{document}
\begin{CJK*}{UTF8}{gbsn}

\title{Individualized Dynamic Latent Factor Model for Multi-resolutional Data with Application to Mobile Health}
\author[1]{Jiuchen Zhang}
\affil[1]{Department of Statistics, University of California, Irvine}
\author[2]{Fei Xue}
\affil[2]{Department of Statistics, Purdue University}
\author[1]{Qi Xu}
\author[3]{Jung-Ah Lee}
\affil[3]{Sue \& Bill Gross School of Nursing, University of California, Irvine}
\author[1]{Annie Qu}
\date{}
\maketitle

\begin{abstract}
Mobile health has emerged as a major success for tracking individual health status, due to the popularity and power of smartphones and wearable devices. This has also brought great challenges in handling heterogeneous, multi-resolution data which arise ubiquitously in mobile health due to irregular multivariate measurements collected from individuals. In this paper, we propose an individualized dynamic latent factor model for irregular multi-resolution time series data to interpolate unsampled measurements of time series with low resolution. One major advantage of the proposed method is the capability to integrate multiple irregular time series and multiple subjects by mapping the multi-resolution data to the latent space. In addition, the proposed individualized dynamic latent factor model is applicable to capturing heterogeneous longitudinal information through individualized dynamic latent factors. Our theory provides a bound on the integrated interpolation error and the convergence rate for B-spline approximation methods.  Both the simulation studies and the application to smartwatch data demonstrate the superior performance of the proposed method compared to existing methods. 
\end{abstract}


\section{Introduction}
\label{sec:introduction}

\begin{figure}[ht]

\centering

\includegraphics[width=.8\linewidth]{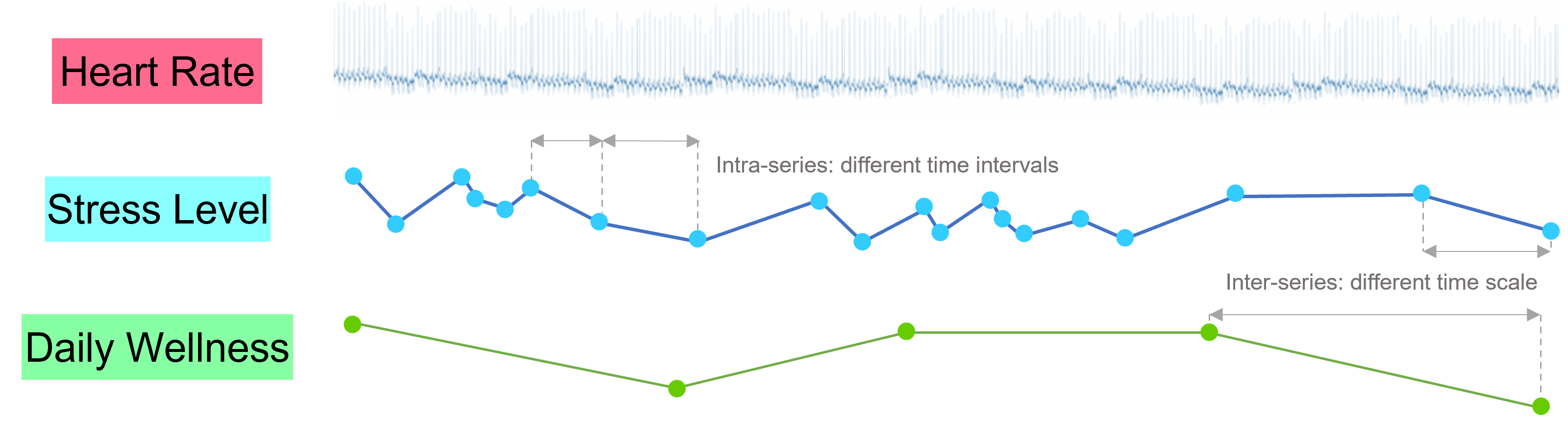}

\caption{Intra-series and inter-series irregularities of Irregular Multi-resolution Time Series in wearable device data.}

\label{fig:ehr}

\end{figure}

With recent developments in technology, mobile health has begun to play an important role in personalized treatment and intervention due to the wide usage of smartphones and wearable devices. A wealth of longitudinal data from wearable devices track people's  physical activities and health status, which enables us to deliver non-invasive interventions in real time. To address the unique challenges presented by this data, including its high heterogeneity, multi-resolution, and non-linearity, we need to develop statistical methods, theories, and computational tools. The data often includes both dense observations over a long period of time and also sparse observations due to a large proportion of missing data. Figure \ref{fig:ehr} illustrates an example of mobile health data for monitoring  heart rate, stress, and daily wellness, where heart rate is measured much more frequently than stress and daily wellness.

 In this paper, we are particularly interested in irregular multi-resolution time series as the data presents three aspects of irregularity \citep{sun2020review}: irregular intra-series due to irregular time intervals within each time series; irregular inter-series due to varying  sampling rates among multivariate time series from the same subject;  and irregular inter-subject measurement variations due to different  time stamps  across different subjects. In addition to irregular and multi-resolution data features, subjects can be highly heterogeneous in terms of demographics, genetic characteristics, medical history, lifestyle, and many unobserved attributes \citep{conway2011analyzing}. Thus, each subject is expected to have a unique trajectory on measurements of interest. Traditional homogeneous models are no longer suitable for this type of data \citep{hamilton2020time, petris2009dynamic, wang2016functional}, so individualized modelling and learning for heterogeneous data are in great demand.

In particular, we are motivated by a stress management study for caregivers of dementia patients which uses mobile health data to intervene with subjects experiencing high stress \citep{lee2021caregiver}. To  administer  intervention for subjects experiencing high stress, we need to capture the trajectories of the subject's physiological information such as heart rate, heart rate variability, physical activities, and daily wellness. Some measurements are low-resolution time series, and therefore interpolating unobserved data from irregular multi-resolution time series could play an important role in performing downstream analyses in prediction,  classification, or clustering \citep{jensen2012mining}.

Traditional polynomial and spline methods can provide interpolation \citep{de1978practical} for a single time series. However, they are not effective for incorporating correlations among time series, which may lead to information loss shared by  multivariate time series from the same subject. Multivariate time series \citep{hamilton2020time} and the dynamic linear model \citep{petris2009dynamic} are capable of handling multiple time series with missing values \citep{gomez1999missing}. However, the high missing rate of low-resolution time series makes it difficult to infer and predict trajectories of longitudinal data  \citep{jones1980maximum}, especially when the occurrence of ultra-sparse time series could lead to degenerated interpolation of missing values due to a large gap from prior observations. In addition, existing approaches require the stationarity assumption which might be difficult to satisfy or verify.

Functional data methods such as functional principal component analysis \citep{wang2016functional, hall2006properties,yao2006penalized} and functional regression \citep{yao2005functional, james2001functional} are useful when analysing longitudinal data. Among these works in functional data analysis, to address the classification of new curves and account for subject heterogeneity, \citet{james2001functional} extend linear discriminant analysis to functional data. They model each predictor as a smooth curve and transforms the curve to a vector of coefficients through basis functions. \citet{james2002generalized} further applies a similar idea to generalized linear models with functional predictors. Additionally, \citet{james2000principal} proposed a principal component method for irregular and sparse data. These existing methods are designed for a single outcome. For instance, \citet{james2002generalized} modelled each time series separately when dealing with multiple time series in prediction models, which ignored the potential correlations among multivariate time series. 

To the best of our knowledge, only a limited literature addresses multivariate functional data analysis \citep{jacques2014model,berrendero2011principal,chiou2014linear,chiou2016pairwise}, and even fewer consider the case of irregular observed data \citep{happ2018multivariate}. Specifically, \citet{happ2018multivariate} introduced a functional principal component analysis  tailored for multivariate functional data with varying dimensions. Although the covariance between time series are considered in multivariate functional data analysis, they generally do not address heterogeneity from different subjects. \citet{volkmann2023multivariate} proposed an additive mixed-effect model for multivariate functional data, where random effects incorporate subject heterogeneity for each subgroup. However, when subject trajectories do not exhibit grouping structures, the mixed-effects model has a limitation.

The interactive fixed-effect models also incorporate subject heterogeneity in longitudinal data/time series data through interactions between individual effect factors and time-effect factors \citep{bai2009panel, bonhomme2015grouped, athey2021matrix}. Interactive fixed-effect models are mainly applied as an  extension to linear functional regression models, whereas our goal is to interpolate missing values utilizing multiple time series. Furthermore, their approaches account for the interaction between individual effects and time effects for each time series. In contrast, we map multivariate time series onto a shared latent space and directly estimate individualized dynamic latent factors without additional decomposition steps.

State-of-the-art deep learning methods are widely used for supervised and unsupervised learning for performing both interpolation and prediction tasks \citep{sun2020review}. For example, recurrent neural networks \citep{hochreiter1997long} are powerful for sequential data. However, due to their complex architecture and the large number of parameters involved,  recurrent neural networks require massive training data to guarantee good performance. Furthermore, recurrent neural networks require a homogeneous assumption for subjects,  making them unsuitable for individualized predictions on trajectories, especially when the sample size of training data is limited.

We propose an individualized dynamic latent factor model to integrate multivariate  data from heterogeneous subjects. The proposed method incorporates irregular multi-resolution time series from each subject utilizing individual-wise dynamic latent factors,  in addition to integrating population-wise information via shared latent factors across different subjects. Specifically, we estimate the dynamic latent factors through a nonparametric model such as  B-spline approximation and establish the corresponding algorithm based on the alternating gradient descent \citep{tseng2009coordinate}. In addition, we extend the dynamic latent factor model to a more general nonparametric framework beyond the B-spline approximation and establish consistency of the proposed interpolation model.

The proposed method has the following advantages. First, through mapping observed irregular time series to the unobserved latent space, the dynamic latent factor model allows us to effectively utilize the multi-resolution time series since the trajectories of correlated multiple time series information can be borrowed from each other through shared latent space. Consequently, the proposed interpolation for the missing data is more precise compared to interpolation from a single time series.

Second, our method integrates data not only from multivariate time series but also across multiple subjects. Through characterizing a population-wise association between dynamic latent factors and observed time series, the latent factors shared across subjects allow us to capture homogeneous features in addition to heterogeneous features. Thus, the proposed individualized  dynamic latent factor model model aggregates time series from all subjects to interpolate missing data, which can make a significant improvement in interpolation especially when the resolution of a time series of interest is  sparse.

Third, the proposed individualized  dynamic latent factor model is applicable for time series with a complex trajectory. In particular,  in contrast to stationary or Markov chain assumptions required by multivariate time series \citep{hamilton2020time} and the dynamic linear model \citep{petris2009dynamic}, we only  require  a smoothness assumption \citep{claeskens2009asymptotic} if B-spline approximation is implemented in the dynamic latent factor modeling. Therefore, the proposed method can model non-stationary processes or time series with abrupt changes, which is particularly useful in practice as abrupt changes in time series data  can often occur.

\section{The Proposed Method}
\label{sec:methodology}
\subsection{General Methodology}

In this subsection, we propose an individualized dynamic latent factor model to capture the trajectory of multi-resolution time series while preserving time-invariant shared information across subjects for each time series. 

\begin{figure}[ht]
\centering
\includegraphics[width=\linewidth]{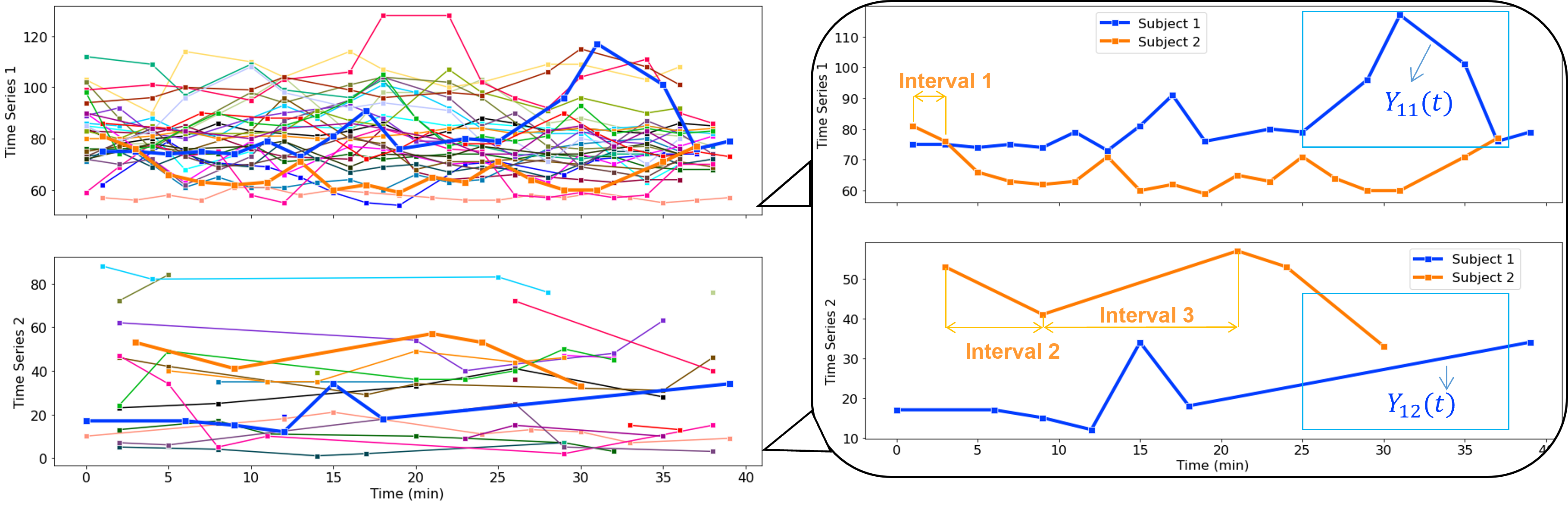}
\caption{Irregular multi-resolution time series with multi-resolution and irregular time intervals. The left plot provides two time series obtained from the smartwatch data where different colours represent different caregivers. The right figure shows two subjects from these two time series.}
\label{fig:multiresolution1}
\end{figure}

We consider a $J$-dimensional multivariate time series for the subject $i$, where $i = 1,\cdots, I$, 

$${Y}_i(t) = (Y_{i1}(t), Y_{i2}(t), ..., Y_{iJ}(t) )^T,$$
where $Y_{ij}(t)$ denotes the $j$-th time series for $t \in [0,T]$, a finite interval. For each time series $Y_{ij}(t)$, there are $K_{ij}$ observations at time points in $\mathbb{T}_{ij} = \{t_{ijk} | k=1,\cdots,K_{ij}, t_{ijk} \in [0,T] \}$. We illustrate observations of the $J=2$ time series in Figure \ref{fig:multiresolution1}, where different colours represent different subjects.

The irregularity of multivariate time series imposes great challenges in that the number of observations $K_{ij}$ could be different for different subjects and time series due to the multi-resolution nature. Specifically, for any pair of time series $Y_{ij}(t)$ and $Y_{ij^{\prime}}(t)$, the time interval $t_{ijk} - t_{ij(k-1)}$ could be different from $t_{ij^{\prime}k} - t_{ij^{\prime}(k-1)}$, as illustrated in interval $1$ and interval $2$ of Figure \ref{fig:multiresolution1}. Similarly, within a single time series $Y_{ij}(t)$, time intervals $t_{ijk} - t_{ij(k-1)}$ and $t_{ijk^{\prime}} - t_{ij(k^{\prime}-1)}$ could also be different, as shown in interval $2$ and interval $3$ of Figure \ref{fig:multiresolution1}. Furthermore, the sets of time points $\mathbb{T}_{ij}$ may also differ between subjects. That is, even if the resolution or sample rate for each time series $j$ is the same for all subjects, the time points $t_{ijk}$ and $t_{i^{\prime}jk}$ may not be the same.

One of our goals is to interpolate unsampled points for a low-resolution time series. Without loss of generality, let the $J$-th time series $Y_{iJ}(t)$ be the time series of interest. Due to the low resolution of the  time series, values of the time series $Y_{iJ}(t)$ at some time points $\{t \in \mathbb{T}_{ij}|j = 1, \cdots, J-1\}$ might not be observed. For illustration, on the right side of Figure \ref{fig:multiresolution1}, there are observations in the blue box for the time series $Y_{11}(t)$, while time series $Y_{12}(t)$ is not observed. In addition, on the left side of Figure \ref{fig:multiresolution1}, we observe that time series with one particular lower resolution $Y_{i2}(t)$ have far fewer observed time points than other time series, yet contain observed time points the other series $Y_{i1}(t)$ do not have.

In the following, we propose to model each time series $Y_{ij}(t)$ for $i = 1, \cdots, I$, $j = 1, \cdots, J$, $t \in [0,T]$ by
\begin{equation}
\label{eq:DLF0}
    Y_{ij}(t) = {f}_j^T {\theta}_i(t) + \epsilon_{ij}(t),
\end{equation}
where ${f}_j \in \mathbb{R}^R$ is a vector of population-wise latent factors corresponding to the $j$-th time series, and $R$ is the dimension of the latent space; the dynamic latent factor ${\theta}_i(t) = (\theta_{i1} (t),..., \theta_{iR}(t))^{T}$ is a vector of continuous functions of $t$ for subject $i$ capturing individual-specific features; and the random noises $\epsilon_{ij}(t)$ are independent and identically distributed. We let ${F} = ({f}_1,\cdots,{f}_J )^T \in \mathbb{R}^{J \times R}$ denote the latent factor matrix. 

By modelling each time series through the inner product of latent factors ${f}_j$ and ${\theta}_i(t)$ in (\ref{eq:DLF0}), we are able to integrate data from multi-resolution time series and different subjects. In addition, mapping multivariate time series to latent space via dynamic latent factor ${\theta}_i(t)$ allows us to utilize information from multi-resolution time series. On the other hand, we require latent factor ${F}$ to be shared among subjects so information across subjects can be borrowed. The data integration can help us to improve the estimation accuracy of ${F}$ and ${\theta}_i(t)$ and thus the accurate interpolation of the target time series $Y_{iJ}(t)$ can be achieved. Specifically, we utilize the latent factor ${f}_j$ to capture time-invariant features of the $j$-th time series, while using the dynamic latent factor ${\theta}_i(t)$ for time-varying features. 

In contrast to $F$, the dynamic latent factor ${\theta}_i(t)$ represents an individualized time trajectory that can be heterogeneous for different individuals. In most longitudinal data, such as our mobile health study, different time series could be correlated for the same subject, therefore we can use a few latent factors to represent time-varying individual features. Thus using the existing homogeneous models \citep{goodfellow2016deep, petris2009dynamic}, including the recurrent neural network model, may lead to high interpolation error, as these models are not suitable for learning individualized trajectories.

Additionally, the dynamic latent factor $\theta_i(t)$ for the $i$-th subject allows for the incorporation of multi-resolution time series $Y_{ij}$'s for $j = 1,\cdots, J$, whereas traditional interpolation methods, such as polynomial interpolation and spline interpolation \citep{de1978practical}, only use data from a single time series $Y_{iJ}(t)$. As we mentioned earlier, one of the interests in the time series study is to provide interpolation of low-resolution time series, which may be necessary due to budget or technical limitations in obtaining high-resolution data. Borrowing information from other time series of the same subject is robust for interpolation, especially in the case of low-resolution time series in multi-resolution data. This is because certain variations in the time series may not be captured by low-resolution observations. For example, in Figure \ref{fig:multiresolution1}, the time series $Y_{i2}(t)$ is of interest. As highlighted in blue boxes, the observations of time series $Y_{11}(t)$ suggest that time series ${Y}_1(t)$ changes abruptly and manages to return back to the previous trend. However, the trajectory of time series $Y_{12}(t)$ might miss the abrupt change due to its low resolution, leading to high interpolation errors based only on observed data of $Y_{12}(t)$. In contrast to traditional interpolation methods, the proposed method is able to preserve the abrupt change through estimating ${\theta}_1(t)$ using additional time series $Y_{11}(t)$ and therefore provides more precise interpolation for $Y_{12}(t)$. 
Our method can be broadly applicable to time series with more complex trajectories, such as non-stationary processes or sparse time series. This is particularly useful when prior knowledge of time series is unknown, or when the data patterns indicate that  stationary or non-stationary model assumptions \citep{hamilton2020time, petris2009dynamic} are not satisfied with the data. In practice, the stationary assumption could be too restrictive  \citep{hamilton2020time}. The non-stationary random processes assumption \citep{petris2009dynamic} could also be restrictive, as it typically requires that the state process be a Markov chain.

\subsection{Latent Factor Estimation}

In this subsection, we propose to estimate the dynamic latent factors ${\theta}_i(t)$ using B-spline functions to capture non-linear function patterns. Specifically, we estimate the latent factors ${F}$ and parameters associated with the B-spline by minimizing a regularized square loss on $Y_{ij}(t)$.

We assume that each dynamic latent factor element $\theta_{ir} (t)$ is a function in the Sobolev space $W_{q}^{\alpha}[0, T]$ equipped with a finite $L_{q}$ norm, where $\alpha$ is a smooth parameter such that $\theta_{ir}(t)$ and its weak derivatives up to order $\alpha$ have a finite $L_q$ norm. We approximate $\theta_{ir} (t)$ by a linear combination of B-spline basis functions of order $\alpha+1$, that is,

\begin{equation*}
    \theta_{ir} (t) \approx \sum_{m = 1}^{M} w_{irm} B_m(t), \quad r = 1,\cdots,R,
\end{equation*}
where $B_m$ ($m=1,2,\cdots, M$) are basis functions of smoothing degree $\alpha$, and ${W} = \{w_{irm} \} \in \mathbb{R}^{I \times R \times M}$ consists of weights for each basis function $B_m$. Specifically, on the time interval $[0, T]$, we use a sequence of $a$ interior knots $0 < \kappa_1 < \kappa_2 < \cdots < \kappa_a < T$, and therefore the number of basis functions $M = a+\alpha+1$. 

In the context of time series or longitudinal data, the splines method is effective in modelling non-linear trends over time \citep{welham2009smoothing} and is also flexible for modelling correlated longitudinal data \citep{de1978practical}. To model irregular multi-resolution time series data with correlations among time points and multiple time series within the same subject, we estimate the parameters ${F}$ and ${W}$ by minimizing the following square loss on  $Y_{ij}(t)$ with a $L_2$ penalty \citep{bi2017group, agarwal2009regression, salakhutdinov2007restricted}:
\begin{equation}
\label{eq:loss}
L({F},{W}) = \sum_{i=1}^I \sum_{j=1}^J \sum_{t \in \mathbb{T}_{ij}} \{Y_{ij}(t) - {f}_j^T {W}_i {B}(t) \}^2 + \lambda (\|{F}\|_F^2 + \|{W}\|_F^2),
\end{equation}
where ${W}_i = \{w_{irm}\} \in \mathbb{R}^{R \times M}$ and ${B}(t) = \{B_{m}(t)\} \in \mathbb{R}^M$, $\lambda$ is the  tuning parameter, and $\| \cdot \|_F$ denotes the Frobenius norm. We estimate ${F}$ and ${W}$ through

\begin{equation*}
(\widehat{{F}}, \widehat{{W}}) = \argmin_{{F},{W}} L({F},{W}).
\end{equation*}

We use the Frobenius-norm penalty to control the non-smoothness of the fitted curve  by filtering out spurious coefficients of the latent factor matrix ${F}$ and spline tensor ${W}$. The regularization of ${W}$ enables us to use a relatively large number of interior knots without knowing the number of knots while shrinking some spline coefficients towards zero. Allowing more interior knots leads to more flexibility in modelling the non-linear trajectory \citep{claeskens2009asymptotic}.

We can also choose other approximation methods for dynamic latent factors ${\theta}_i(t)$, such as the kernel approach \citep{wenzel2021novel} or deep learning methods \citep{goodfellow2016deep}. However, the interpolation accuracy is influenced by the approximation error of the dynamic latent factors, which is  determined by the choice of approximation methods. 

Once the estimators $\widehat{{F}}$ and $\widehat{{W}}$ are obtained, the proposed interpolation at any time $t \in [0,T]$ is calculated by

\begin{equation}
\label{eq:DLF1}
    \widehat{Y}_{ij}(t) = \widehat{{f}}_j^T \widehat{{W}}_i {B}(t).
\end{equation}

Note that equation (\ref{eq:DLF1}) provides a general formula for all time series at any time point in the range $[0, T]$. However, in practice, we might only be interested in interpolating a single time series $\widehat{Y}_{iJ}(t)$ for $\{t \in \mathbb{T}_{ij}|j = 1, \cdots, J-1\}$.

\section{Theory}
\label{sec:theory}
In this section, we develop the theoretical properties of the proposed method based on a sample estimator from a Sobolev space in addition to providing the theoretical properties of the estimation with the B-spline approximation. Specifically, we establish the asymptotic property of the integrated interpolation error and provide the rate of convergence for the proposed estimator when the parameter space is a Sobolev space. Finally, we provide a concrete convergence analysis for B-spline approximation to demonstrate the theoretical properties of our implemented model in Section \ref{sec:methodology}.

We first consider a general result for $\theta_{ir}(t) \in W_{q}^{\alpha}[0, T]$ for $\alpha > 1$ and $q \ge 2$, where $W_{q}^{\alpha}[0, T]$ is a Sobolev space with a finite $L_{q}$ norm. The parameter $\alpha$ is a smooth parameter such that $\theta_{ir}(t)$ and its weak derivatives up to order $\alpha$ have a finite $L_q$ norm. Additionally, we assume that $K_{ij} \sim K$ for some $K$, where $a \sim b$ when $a$ and $b$ have the same order.

Since our primary goal is the interpolation of ${Y}(t)$, we focus on the convergence property of the interpolation values instead of the latent factor recovery. Suppose the time series
$$
    Y_{ij}(t) = \psi_{ij}(t) + \epsilon_{ij}(t),
$$
where $\psi_{ij}(t) = \sum_{r=1}^R f_{jr}\theta_{ir}(t)$, ${\Psi}(t) = \left(\psi_{ij}(t)\right)$ is a vector of $\psi_{ij}(t)$ for $i = 1,\cdots, I$ and $j = 1, \cdots, J$, and the random noises $\epsilon_{ij}(t)$ are independently and identically distributed with mean $0$ and variance $\sigma^2$. For simplicity, we write ${\Psi}(t)$ as ${\Psi}$ in this paper. By $\theta_{ir}(t) \in W_{q}^{\alpha}[0, T]$, we have $\psi_{ij}(t) \in W_{q}^{\alpha}[0, T]$ by construction. For time series  $Y_{ij}(t)$ of ${Y}(t)$, we define the $L_2$-loss function  as
$$
 l\left({\Psi}, Y_{ij}(t)\right)=\left\{Y_{ij}(t)-\psi_{ij}(t)\right\}^{2}.
$$

 Let $\Omega$ be the set of observations, $|\Omega| = \sum_{ij}K_{ij}$ be the number of observations and $J({\Psi})$ be a non-negative penalty function. For example, we have $J({\Psi}) =\sum_{ij} \{\int_{0}^T|\psi_{ij}^{(\alpha)}(t)|^qdt \}^{1/q}$ since $\psi_{ij}(t) \in W_{q}^{\alpha}[0, T]$. Then the overall object function is
\begin{equation}
\label{eq:losspf}
 L({\Psi} | {Y})=\frac{1}{|\Omega|}\sum_{\left(i, j, t\right) \in \Omega} l\left({\Psi}, y_{ij}(t)\right)+\lambda_{|\Omega|} J({\Psi}),
\end{equation}
where $\lambda_{|\Omega|}$ is a tuning parameter for the penalization. To establish the convergence rate, we introduce the following assumption:
\begin{assumption}
\label{asm:edf}
For the empirical distribution $Q_{ij,n}$ of $t_{ij1}, \ldots, t_{ijK_{ij}}$, 
$$
Q_{ij,n}(t) = \frac{1}{K_{ij}}\sum_{k=1}^{K_{ij}} \mathbf{1}_{t_{ijk} < t},
$$
where $\mathbf{1}_A$ is the indicator function of event $A$, there exists a distribution function $Q_{ij}(t)$ with positive continuous density such that, 
$$\sup _{t \in[0, T]}\left|Q_{ij,n}(t)-Q_{ij}(t)\right|=o\left(K_{ij}^{-1}\right).$$
\end{assumption}

Assumption \ref{asm:edf} assumes that the empirical distributions of observed time points converge to a distribution of $t$ with positive continuous density, which is typically the uniform distribution for random samples. Such an assumption is common in spline approximation. Similarly, in kernel approximation, the observed points are  assumed to be asymptotically uniformly distributed \citep{wenzel2021novel}. Note that when the observed time points $t_{ijk}$ are random and sampled from the true distribution $Q_{ij}(t)$, Assumption \ref{asm:edf} is satisfied naturally by the Glivenko–Cantelli theorem \citep{sharipov2011glivenko}.

Let ${\Psi}_0$ be the true parameters and $\mathcal{S} = \left\{{\Psi}:\psi_{ij}(t) \right.$ $= \sum_r f_{jr}\theta_{ir}(t)$, $\|{F} \|_{\infty} \le c_0, $ $\left.\theta_{ir}(t) \in W_{q}^{\alpha}[0, T] \; \text{for} \; i = 1,\cdots, I,j = 1, \cdots, J    \right\}$ be the parameter space which depends on a positive constant $c_0$. We denote ${\hat \Psi}_{|\Omega|}$ as the sample estimator of ${\Psi}_0$  satisfying:
\begin{equation}
\label{eq:assumptheta}
L({\hat \Psi}_{|\Omega|} | {Y}) \le \displaystyle \inf_{\Psi \in \mathcal{S}} L({\Psi} | {Y})+\tau_{|\Omega|},
\end{equation}
where $\displaystyle \lim_{|\Omega| \rightarrow \infty} \tau_{|\Omega|} = 0$. This condition implies that ${\hat \Psi}_{|\Omega|}$ is close to the global minimizer of  $ L({\Psi} | {Y})$ when $|\Omega| \rightarrow \infty$. This necessity arises because, in practice, obtaining the exact global minimum is often impractical due to the non-convex nature of the function $L$. 

In addition to the above assumptions, Assumptions \ref{asm:epsilon} and \ref{asm:designmatrix} are provided in the Supplementary
Materials and are standard assumptions for nonparametric approximations on the basis functions,
 observation distribution, and random noises.

We establish the convergence of ${\hat \Psi}_{|\Omega|}$ in the following Theorem \ref{thm:convgen}.

\begin{theorem}
\label{thm:convgen}
Suppose ${\hat \Psi}_{|\Omega|}$ is a sample estimator satisfying \eqref{eq:assumptheta}. Then for $K_{ij} \sim K$ for some order $K$, and under Assumptions \ref{asm:edf}, \ref{asm:epsilon}, and \ref{asm:designmatrix}, we have:
\begin{equation}
\label{eq:errorgen}
    \sqrt{\frac{1}{IJ} \sum_{i=1}^{I} \sum_{j = 1}^{J} \int_{0}^{T}\left\{\hat \psi_{ij}(t)-\psi_{0,ij}(t)\right\}^2dQ_{ij}(t)} = O_{p}\{(IJK)^{-\alpha/(2\alpha+1) }\} + o_{p}(K^{-1/2}),
\end{equation}
when $\tau_{|\Omega|} = o((IJK)^{-2\alpha/(2\alpha+1)})$ and 
$\lambda_{|\Omega|} \sim (IJK)^{-2\alpha/(2\alpha+1)}$.
\end{theorem}

Theorem \ref{thm:convgen} indicates that the average interpolation error in terms of integrated squared loss with respect to time  converges to zero when $K$ and $IJ$ go to infinity. The first term of the error bound in \eqref{eq:errorgen} is due to the approximation bias of the dynamic latent factor which is determined by the smoothness of $\theta_{ir}(t)$. This approximation bias becomes negligible when the total number of observations goes to infinity. The second term of the error bound is determined by the difference between the empirical distribution and the reference distribution $Q_{ij}(t)$, which converges to zero when the number of observations of each time series $K_{ij}$ goes to infinity, according to Assumption \ref{asm:edf}.

Theorem \ref{thm:convgen} demonstrates the benefits of integrating information across subjects and time series. The interpolation ${\hat \Psi}_{|\Omega|}$ converges faster to the true value ${\Psi}_{0}$ if the number of observed subjects $I$ or the number of observed time series $J$ are larger. That is, when we integrate more time series and more subjects, we obtain better interpolation. 

Next, we extend Theorem \ref{thm:convgen}  to the case when the B-spline approximation is applied, that is,
the parameter space becomes $\mathcal{S}_{M} = \left\{{\Phi}:\phi_{ij}(t)  \right.$ $= \sum_{r,m}f_{jr}w_{irm}B_{m}(t)$, $\left.\|{F} \|_{\infty}, \|{W} \|_{\infty} \le c_0    \right\}$ where ${F} = (f_{jr})$, ${W} = (w_{irm})$ and $M$ is the number of basis functions. Similar to previous notations, we define the $L_2$-loss function  as
$
 l\left({\Phi}, Y_{ij}(t)\right)=\left\{Y_{ij}(t)-\phi_{ij}(t)\right\}^{2}.
$ and the overall object function as
$
 L({\Psi} | {Y})=\sum_{\left(i, j, t\right) \in \Omega} l\left({\Psi}, y_{ij}(t)\right)/|\Omega|+\lambda_{|\Omega|} J({\Psi}).
$
Additionally, we let the penalty be the $L_2$ penalty defined in Section \ref{sec:methodology}, that is, $J({\Phi}) =\|{F}\|_2^2+\|{W}\|_2^2$. 

We also denote ${\hat \Phi}_{|\Omega|} = (\hat \phi_{ij}(t))$ as the sample estimator of ${\Psi}_0$,  satisfying:
\begin{equation}
\label{eq:assumpthetanew}
L({\hat \Phi}_{|\Omega|} | {Y}) \le \displaystyle \inf_{\Phi \in \mathcal{S}_M} L({\Phi} | {Y})+\tau_{|\Omega|},
\end{equation}
where $\displaystyle \lim_{|\Omega| \rightarrow \infty} \tau_{|\Omega|} = 0$.

We establish the asymptotic property of ${\hat \Phi}_{|\Omega|}$ in the following theorem.

\begin{theorem}
\label{thm:convB}
Let ${\hat \Phi}_{|\Omega|}$ be a sample estimator satisfying \eqref{eq:assumpthetanew}.  Then for $K_{ij} \sim K$ for some order $K$, and under Assumptions \ref{asm:edf}, \ref{asm:epsilon}, and \ref{asm:designmatrix}, we have:
$$
\sqrt{\frac{1}{IJ} \sum_{i=1}^{I} \sum_{j = 1}^{J} \int_{0}^{T}\left\{\hat \phi_{ij}(t)-\psi_{0,ij}(t)\right\}^2dQ_{ij}(t)} = O_{p}\{(IJK)^{-\alpha/(2\alpha+1) }\} + o_{p}(K^{-1/2}),
$$
when $\tau_{|\Omega|} = o((IJK)^{-2\alpha/(2\alpha+1)})$ and 
$\lambda_{|\Omega|} \sim (IJK)^{-2\alpha/(2\alpha+1)}$.
\end{theorem}

Theorem \ref{thm:convB} shows that the convergence rate of the proposed estimator in  Section \ref{sec:methodology} is $O_{p}\{(IJK)^{-\alpha/(2\alpha+1)}\} + o_{p}(K^{-1/2})$. Note that to obtain the convergence rate we require that the penalty parameter $\lambda$ shrink to zero at a rate of $(IJK)^{-2\alpha/(2\alpha+1)}$.

\section{Computation}
\label{sec:computation}

In this section, we introduce an algorithm and implementation details of the proposed method. Specifically, we  utilize the alternating gradient descent (A-GD) algorithm \citep{tseng2009coordinate} to estimate latent factors ${F}$ and ${W}$.

The alternating gradient descent (A-GD) algorithm provided in Algorithm 1 is a generalization of the block coordinate gradient descent method \citep{tseng2009coordinate}, which is especially useful in matrix decomposition and tensor decomposition \citep{zhao2015nonconvex,zhang2022tensor,bi2018multilayer}. The main idea of the algorithm is to iteratively update each ${F}$ and ${W}_i$ for $i=1,\cdots,I$, while keeping the others fixed. The advantage of this algorithm is that the latent factor matrices naturally provide a block structure of the parameters, and updating ${F}$ and ${W}_i$ enables us to transform the non-convex optimization to a convex optimization. In addition, it can further decrease the number of iterations and lead to faster convergence compared with the gradient descent algorithm. This is because we can use a larger step size when updating blocks of parameters instead of entire parameters \citep{jain2013low}. 

Specifically, let ${F}^{(s)}$ and ${W}^{(s)}_i$ denote the estimated ${F}$ and ${W}_i$ at the $s$-th iteration, and let $L^{(s)} = L({F}^{(s)},{W}^{(s)})$ denote the corresponding loss. We update each ${F}^{(s-1)}$ and ${W}_i^{(s-1)}$ along the direction of the partial derivatives $
\partial L({F},{W}^{(s-1)})/\partial {F}
$ and 
$
\partial L({F}^{(s-1)},{W}_i)/\partial {W}_i
$ at each iteration.

\noindent $\overline{\mbox{\underline{\makebox[\linewidth]{\texttt{Algorithm 1:} Alternating Gradient Descent}}}}$

\begin{enumerate}[]

\item \texttt{Initialization} Set stopping error $\epsilon$, rank $R$, tuning parameter $\lambda$, step size $\alpha$, the basis functions $B_m$ ($m = 1,\cdots, M$) and initial values ${F}^{(0)}$ and ${W}^{(0)}$. \\

\item \texttt{Latent factor update}  At the $s$-th iteration ($s \ge 1$):

(i) Update ${F}^{(s)}$: ${F}^{(s)} \leftarrow {F}^{(s-1)} - \alpha \frac{\partial L({F},{W}^{(s-1)})}{\partial {F}}$.

(ii) Update each ${W}^{(s)}_i$: ${W}^{(s)}_i \leftarrow {W}^{(s-1)}_i - \alpha \frac{\partial L({F}^{(s-1)},{W}_i)}{\partial {W}_i}$. \\

\item  Stop if $\frac{|L^{(s+1)} - L^{(s)}|}{ L^{(s)}} < \epsilon$.

\end{enumerate}
\noindent\makebox[\linewidth]{\rule{\linewidth}{0.4pt}}

To select rank $R$, tuning parameter $\lambda$, and step size $\alpha$, we conduct a grid search through minimizing the mean square error on the validation set. Our empirical study shows that the tuning parameter $\lambda$ is quite robust and would not change the numerical performance much compared to rank $R$ and stepsize $\alpha$. Thus, to save computational cost, we tune the $\lambda$ first, and conduct a grid search on pairs of rank $R$ and step size $\alpha$ after. The results from cross-validation simulations, as well as the performance of the proposed method across different rank values, can be found in the Supplementary Materials.

In addition, selection of basis functions and determining the number of knots are important here. Based on the practice of B-spline approximation, we require the number of basis functions $M$ to be large enough so that there is at least one observation in each interval. However, in practice, even if this assumption is mildly violated, we can still obtain a reasonable interpolation accuracy due to the penalty term. In our numerical study, we utilize the evenly spaced $a$ knots which are smaller or equal to the smallest $K_{ij}$ for $i = 1, \cdots, I$ and $j=1, \cdots, J$. This allows us to model the trajectory sufficiently well by utilizing a relatively large number of knots.

The $L_2$ penalty in equation (\ref{eq:loss}) is selected to avoid over-fitting and scale indeterminacy, which balances computational complexity and model complexity \citep{acar2011scalable}. We can also consider penalty functions used in penalized spline functions, e.g., the integrated squared $q$th-order derivative used in the spline function \citep{claeskens2009asymptotic} or the total variation penalty used in \cite{jhong2017penalized}. However, based on our simulation studies, utilizing these penalty terms results in similar interpolation accuracy as the $L_2$ penalty after proper hyper-parameter tuning. 

\section{Simulations}
\label{sec:simulation}

\subsection{General Setting}
\label{sec:simulation0}
In this section, we conduct simulations to investigate the empirical performance of the proposed individualized dynamic latent factor model and compare it with existing methods under six different settings. Specifically, we compare the proposed method with six competing methods; namely, the smoothing spline \citep{de1978practical}, multivariate time series \citep{hamilton2020time}, dynamic linear model \citep{petris2009dynamic}, functional principal component analysis \citep{wang2016functional}, recurrent neural network and deep recurrent neural network \citep{hochreiter1997long}. 

Here, the smoothing spline applies to the $J$-th time series for each subject separately with smoothing degree $k=3$, which is implemented by the Python package \texttt{scipy.interpolate} \citep{de1978practical}, where knots are selected by the function \texttt{UnivariateSpline} automatically. The  multivariate time series and dynamic linear model implement multivariate time series for each subject. To deal with irregular time intervals, unobserved points are treated as missing values. The  multivariate time series and dynamic linear model are implemented by the Python package \texttt{statsmodels.tsa} \citep{hamilton2020time} and \texttt{pyro} \citep{petris2009dynamic}. The functional principal component analysis integrates the $J$-th time series for all subjects together. To handle irregular time intervals, we apply the functional principal component analysis through B-spline functional basis using the Python package \texttt{scikit-fda} \citep{wang2016functional}. We implemented two recurrent neural network models using the Python package \texttt{tensorflow} \citep{hochreiter1997long} using masking layers to handle irregularity, where the recurrent neural network model contains one recurrent neural network layer with 32 units and one dense layer, and the deep recurrent neural network model contains the same three recurrent neural network layers with 32 units and one dense layer. Both recurrent neural network models are trained by the Adam optimizer \citep{kingma2014adam} with 20 epochs. For the individualized  dynamic latent factor model, we utilize evenly spaced internal knots with the number of basis function $M = 300$, the smoothing degree $p = 3$.

We generate simulated data according to equation (\ref{eq:DLF0}) with ${f}_j \sim N(0, {I}_R)$ and $\epsilon_{ij}(t) \sim N(0,0.5^2)$. In each setting, we let the latent space of dimension, the number of subjects, the number of time series, and the time range be $R = 3$, $I = 30$, $J = 5$, and $T = 1000$, respectively. In each setting, we only generate time points $t=1,\cdots, T$ as most of the competing methods cannot handle continuous time points $t \in [0,T]$, while the proposed method is able to handle them.

We assess interpolation performance by examining the mean square error over the training and testing set based on 50 replications. Specifically, we are interested in the interpolation of the $J$-th time series. Thus, we calculate the mean square error over training and testing sets of the $J$-th time series, that is, $\sum_{i=1}^I \sum_{t \in \mathbb{T}_{iJ}} \{Y_{iJ}(t) - \hat Y_{iJ}(t)\}^2/\sum_{i=1}^I|\mathbb{T}_{iJ}|$ and $\sum_{i=1}^I \sum_{t \not\in \mathbb{T}_{iJ}} \{Y_{iJ}(t) - \hat Y_{iJ}(t)\}^2/\{ IT - \sum_{i=1}^I|\mathbb{T}_{iJ}|\}$, where $|\mathbb{T}_{iJ}|$ denotes the number of observations for the time series $Y_{iJ}$. 

\subsection{Multi-resolution Time series}
\label{sec:simulation1}

In this subsection, we investigate interpolation performance for multi-resolution time series under three settings with observed time points at $\mathbb{T}_{ij}$ for $i = 1,\cdots, I$ and $j = 1,\cdots,J$.

In each setting, we let the dynamic latent factors be 
$$
\begin{aligned}
\theta_{i1} (t) &= 2 \exp\{-\frac{(t-60-10i)^2}{50}\}+4 \exp\{-\frac{(t-70-10i)^2}{20}\},  \\
\theta_{i2} (t) &= i \times 0.02\log(t+1),  \\
\theta_{i3} (t) &= \cos(0.12 \pi t + 1),
\end{aligned}
$$
where $\theta_{i1} $ represents two pulses at time points $60+10i$ and $70+10i$ for $i = 1,\cdots,I$; $\theta_{i2}$ represents a time trend that varies among subjects, and $\theta_{i3}$ represents a seasonal trend. 

To evaluate the performance of the proposed method under different observation processes, we consider three settings, where the numbers of observations are similar. Specifically, the three settings of the observation points of training sets are as follows:

\texttt{Setting 1.1:}
$$
\begin{aligned}
&Y_{i1}, \cdots, Y_{i4}: \mathbb{P}(t \in \mathbb{T}_{ij}) = 0.8 \ \text{for} \  t=1,\cdots,1000; j=1,\cdots,4, \nonumber \\
&Y_{i5}: \mathbb{P}(t \in \mathbb{T}_{i5}) = 0.2 \ \text{for} \  t=1,\cdots,1000. \nonumber
\end{aligned}
$$

\texttt{Setting 1.2:}
$$
\begin{aligned}
&Y_{i1}, \cdots, Y_{i3}: \mathbb{T}_{ij} = \{1, 2, 3\cdots,1000\} \ \text{for} \  j=1,2,3, \nonumber \\
&Y_{i4}: \mathbb{T}_{i4} = \{1,3,5,7,\cdots, 999\}, \nonumber \\
&Y_{i5}: \mathbb{T}_{i5} = \{1,5,9,13,\cdots, 997\}.\nonumber
\end{aligned}
$$

\texttt{Setting 1.3:}
$$
Y_{i1}, \cdots, Y_{i5}: \mathbb{P}(t \in \mathbb{T}_{ij}) = 0.7 \ \text{for} \  t=1,\cdots,1000; j=1,\cdots,5.
$$
Setting 1.1 mimics the most complicated situation of multi-resolution data, where $\mathbb{T}_{ij}$'s are different for different subjects and each time series has an irregular time interval. Setting 1.2 considers the situation where multi-resolution time series have evenly spaced and fixed time points. Setting 1.3 considers the same resolution time series with varying time points $\mathbb{T}_{ij}$ for $i = 1,\cdots, I$ and $j = 1,\cdots,J$. For testing sets of three settings, we use unobserved time points.

Table \ref{tab:sim1} provides the mean square error results under Settings 1.1-1.3. The proposed method has the best performance on testing set under all three settings, with more than $40\%$ improvement of mean square error compared to other methods. The competing methods such as  multivariate time series, dynamic linear model, and functional principal component analysis cannot be applied to time series with multi-resolution or different observation time points for each subject. Thus, only the methods of smoothing spline, recurrent neural network and deep recurrent neural network are compared. As one of the most popular interpolation methods, smoothing spline performs the second best in most settings, except it performs the worst under Setting 1.1 with randomly selected time points for low-resolution time series. In contrast, the proposed method is able to borrow information from other time series from the same subject, especially the one with high resolution, and therefore attains better interpolation accuracy. For recurrent neural network and deep recurrent neural network, the performance varies under different settings. In general, the deep recurrent neural network performs better than the recurrent neural network on training set. However, the deep recurrent neural network performs poorly in interpolation under multi-resolution situations. Overall, the two recurrent neural network models perform the worst as they do not incorporate heterogeneity among subjects. The proposed individualized  dynamic latent factor model performs similarly under the two multi-resolution Settings 1.1 and 1.2. The difference is that, for randomly generated observation time points, the standard deviation of mean square error is higher on the testing set.

\begin{table}
\begin{center}
\caption{The mean square error of the proposed method and competing methods under Settings 1.1-1.3. The standard errors are provided in parentheses. IDLFM: the proposed individualized dynamic latent factor model, SS: smoothing spline, RNN: recurrent neural network, DRNN: deep recurrent neural network}
\label{tab:sim1}
\resizebox{\linewidth}{!}{%
\begin{tabular}{c|cc|cc|cc}
\hline
\hline

MSE   & \multicolumn{2}{c|}{Setting 1.1}                   & \multicolumn{2}{c|}{Setting 1.2}                   & \multicolumn{2}{c}{Setting 1.3}                   \\
\hline
      & Training                 & Testing                   & Training                 & Testing                   & Training                 & Testing                   \\
\hline                 
\begin{tabular}[c]{@{}c@{}}IDLFM\\ (Proposed)\end{tabular}  & {0.209 (0.041)} & {0.415 (0.161)} & 0.212 (0.044) & {0.341 (0.067)} & {0.215 (0.023)} & {0.327 (0.041)} \\
SS    & 0.602 (0.257)          & 2.983 (10.453)         & 0.556 (0.282)          & 0.598 (0.346)          & 0.632 (0.280)          & 0.749 (0.460)          \\
RNN   & 4.186 (6.180)          & 4.399 (6.226)          &        0.350 (0.565)  & 6.786 (6.840)          & 0.894 (0.894)          & 1.259 (1.078)        \\
DRNN  & 1.712 (2.504)          & 5.793 (7.878)          & {0.099 (0.050) }         & 6.610 (6.843)         & 0.604 (0.552)         & 3.915 (4.865) \\
\hline
\hline
\end{tabular}%
}
\end{center}
\end{table}

We also investigate the setting with more time series, where J=101, for each subject, where there are 100 time series as covariates and one time series of interest.  Our numerical study shows that the proposed method attains a lower standard deviation when $J$ is higher as more time series are integrated. The mean square error does not improve much because the convergence rate of the proposed method is related to the number of observation points $K_{ij}$.  Additional simulations also illustrate the robustness of the proposed method under various missing mechanisms, such as Missing Completely at Random, Missing at Random and Missing Not at Random.  The detailed simulation results are provided in Table \ref{tab:sim3} and \ref{tab:missingness} of the Supplementary Materials.

\subsection{Multiple Time Series with Heterogeneity and Non-stationarity}
\label{sec:simulation2}

In this subsection, we focus on time series with the same resolution where all subjects and all time series are observed at the same time points $\mathbb{T}_{ij} = \mathbb{T}_{11}$ for $i = 1,\cdots, I$ and $j = 1,\cdots,J$. We investigate how the heterogeneity of subjects and non-stationarity of time series affect the performance of interpolation in the following three settings. They are the same as the settings in Section \ref{sec:simulation1}, except for $\mathbb{T}_{ij}$ and the dynamic latent factors. The sets $\mathbb{T}_{ij}$ are the same within each setting, which we referred to as $\mathbb{T}$. For the set $\mathbb{T}$, the observation index $t \in \{1,\cdots, T\}$ is selected according to Bernoulli distribution with a probability $0.7$, and the unobserved points are treated as the testing set. This mimics the time series setting where all subjects and time series are observed at the same time points but at unevenly spaced time intervals. The dynamic latent factors for each setting are generated as follows:

For $i=1,2,\cdots,30, t=1,2,\cdots,1000,$

\texttt{Setting 2.1:}
$$
\begin{aligned}
&\theta_{i1} (t) = 2\exp\{-\frac{(t-60)^2}{50}\}+4\exp\{-\frac{(t-70)^2}{20}\},\; \theta_{i2} (t) = 0.2\log(t+1),  \\
&\theta_{i3} (t) = \cos(0.12 \pi t + 1).
\end{aligned}
$$

\texttt{Setting 2.2:}
$$
\begin{aligned}
&\theta_{i1} (t) = 2\exp\{-\frac{(t-60-10i)^2}{50}\}+4\exp\{-\frac{(t-70-10i)^2}{20}\},\;  \theta_{i2} (t) = 0.2\log(t+1), \\
&\theta_{i3} (t) = \cos(0.12 \pi t + 1). 
\end{aligned}
$$

\texttt{Setting 2.3:}
$$
\begin{aligned}
&\theta_{i1} (t) = 2\exp\{-\frac{(t-60-10i)^2}{50}\}+4\exp\{-\frac{(t-70-10i)^2}{20}\},   \\
&\theta_{i2} (t) =  i \times 0.02\log(t+1), \; \theta_{i3} (t) = \cos(0.12 \pi t + 1).
\end{aligned}
$$
In Setting 2.1, the dynamic latent factors are the same for all subjects. In Setting 2.2, we change the locations of two pulses in $\theta_{i1}$ to be different for different subjects and keep $\theta_{i2}$ and $\theta_{i3}$ the same as in Setting 2.1. Additionally, in Setting 2.3, we keep $\theta_{i1}$ and $\theta_{i3}$ the same as in Setting 2.2 and change $\theta_{i2}$ to a function where the time trend varies across subjects, where the stationary assumption is highly violated for subjects with large $i$.

Table \ref{tab:sim2} provides the results of all methods under Settings 2.1-2.3. We observe that the proposed method has the best performance under all three settings, with more than 50\% improvement in mean square error compared to other methods. All methods except the smoothing spline perform best in Setting 2.1 and perform worst in Setting 2.3. This is because the subjects are homogeneous in Setting 2.1, and are highly heterogeneous in Setting 2.3.  The proposed method is the most robust compared to other competing methods due to the advantage of integrating multiple time series and information across subjects. In Setting 2.3, the  multivariate time series and dynamic linear model perform worse than the other two settings, Setting 2.1 and 2.2, since the stationary or Markov chain assumption is violated when the non-linear trend dominates the time series. In contrast, the proposed method can model non-linear trends with dynamic latent factors and provides better interpolation. In addition, the functional principal component analysis and two neural network models fail to deliver accurate interpolation due to heterogeneity for different individuals. The individualized  dynamic latent factor model is capable of integrating time series information from heterogeneous subjects.

\begin{table}
\begin{center}
\caption{The mean square error of the proposed method and competing methods under Settings 2.1-2.3. The standard errors are provided in parentheses. IDLFM: the proposed individualized dynamic latent factor model, SS: smoothing spline, MTS: multivariate time series, DLM: dynamic linear model, FPCA: functional principal component analysis, RNN: recurrent neural network, DRNN: deep recurrent neural network}
\label{tab:sim2}
\resizebox{\linewidth}{!}{%
\begin{tabular}{c|cc|cc|cc}
\hline
\hline

MSE   & \multicolumn{2}{c|}{Setting 2.1}                   & \multicolumn{2}{c|}{Setting 2.2}                   & \multicolumn{2}{c}{Setting 2.3}                   \\
\hline
      & Training                 & Testing                   & Training                 & Testing                   & Training                 & Testing                   \\
\hline
\begin{tabular}[c]{@{}c@{}}IDLFM\\ (Proposed)\end{tabular}  & {0.201 (0.020)} & {0.359 (0.046)} & {0.208 (0.026)} & {0.334 (0.049)} & {0.213 (0.024)} & {0.350 (0.046)} \\
SS    & 0.726 (0.275)          & 0.876 (0.501)          & 0.564 (0.278)          & 0.668 (0.466)          & 0.608 (0.253)          & 0.720 (0.457)          \\
MTS   & 0.652 (0.384)          & 0.761 (0.488)          & 0.607 (0.414)          & 0.722 (0.523)          & 0.773 (0.919)          & 0.919 (1.142)          \\
DLM   & 3.410 (3.575)          & 3.405 (3.629)          & 3.421 (3.821)          & 3.536 (4.036)          & 4.840 (8.506)          & 4.838 (8.340)          \\
FPCA  & 4.429 (4.415)          & 4.501 (4.453)          & 4.516 (4.885)          & 4.546 (4.901)          & 6.665 (6.128)          & 6.762 (6.127)          \\
RNN   & 0.869 (0.550)          & 0.865 (0.569)          & 0.805 (0.665)          & 0.818 (0.704)          & 1.584 (3.416)          & 1.568 (3.324)          \\
DRNN  & 0.783 (0.515)          & 0.784 (0.531)          & 0.744 (0.581)          & 0.756 (0.604)          & 1.350 (2.646)          & 1.333 (2.589)           \\
\hline
\hline
\end{tabular}%
}
\end{center}
\end{table}

\section{Real Data Application}
\label{sec:realdata}

In this section, we apply the proposed method to smartwatch data for caregivers of dementia patients from the Dementia Family Caregiver Study, which was done by \cite{lee2021caregiver} in California. 

Studies show that the under-served caregivers of dementia patients frequently experience physical and emotional distress \citep{anthony1988symptoms}. To better monitor and manage caregivers' stress, \cite{lee2021caregiver} conducted home-based, culturally and language-specific interventions for dementia family caregivers, which included stress self-management and caregiving education. Community health workers offered stress reduction techniques including mindful deep breathing and compassionate listening during up to 4 home visits with the caregiver. Additionally, community health workers also provided caregiving education to improve caregivers' health, well-being, and positive interactions with dementia patients during a one-month study period. The intervention employed is not in an online form, as it is pre-planned and remains unaffected by the collected data. To monitor the caregiver’s physical activities and physiology, caregivers are equipped with Wearable Internet of Things (WIOT) devices, such as the smartwatch, for a month. These wearable devices are used to measure the intervention's impact, offering noticeable changes due to the intervention. There are five time series of measurements from the smartwatch data, which are steps, heart rate, activity level, movement, and stress. Among these five measurements, we are interested in evaluating and predicting the stress level of caregivers, which are highly associated with the other four time-series variables \citep{saykrs1973analysis}. Our goal is to interpolate unobserved stress levels, so we can assess the effects of the intervention more accurately through the investigation.
These measurements are observed with multi-resolution. Specifically, activity level and movement are observed every minute; heart rate is collected every two minutes; stress is measured every three minutes; and steps are counted every 15 minutes. Additionally, due to the heterogeneous nature of individuals, the observed time also varies for different caregivers, leading to irregular time series. 

In our study, we remove the first four days and last ten days of data to ensure data quality, as subjects might not be familiar with the devices and have trouble setting up, or stop wearing the devices at the end of the study. There are 33 caregivers and more than 23,000 time points from day 5 to day 20. The average missing rate for the stress variable is $80\%$, while missing rates for steps, heart rate, activity level, and movement are 97\%, 64\%, 37\%, and 18\%, respectively. The varying missing rates among different time series primarily result from differences in the collection frequency of each variable. Additionally, the measurement techniques for each time series and caregivers' charging habits contribute to the missingness. Since the scales of different time-series measurements are different, we standardize each observed value $Y_{ij}(t)$ by subtracting the mean and dividing by the standard deviation of the corresponding time series $Y_{ij}$.

To train the interpolation models and evaluate the performance of the proposed method, we randomly select 30\% of the observations for stress levels as a testing set and assign the remaining stress observations and observations of the other four time-series measurements as a training set. The interpolation model predictions are evaluated by the mean square error of stress predictions on the training and testing sets based on 50 replications.

We compare the proposed individualized  dynamic latent factor model method with the three competing methods, including smoothing spline, recurrent neural network, and deep recurrent neural network, as described in Section \ref{sec:simulation}. Table \ref{tab:real} provides the average mean square error based on both training and testing sets. The proposed individualized  dynamic latent factor model method outperforms the other methods significantly in terms of achieving the lowest mean square error. Specifically, the proposed method reduces 24\%, 61\%, and 72\% of the mean square errors of smoothing spline, recurrent neural network, and deep recurrent neural network methods on testing sets, respectively, and achieves the smallest standard error among all competing methods. This is because the smoothing spline does not fully utilize the other four time-series observations. The recurrent neural network using one layer neural network cannot accommodate complex time-series data. Although the three-layer deep recurrent neural network model performs better than the recurrent neural network model, it still produces much larger mean square error than the proposed individualized  dynamic latent factor model, as the deep recurrent neural network is not capable of handling low sample-size heterogeneous data. 

\begin{table}
\begin{center}
\caption{The mean square error of the proposed method and competing methods for smartwatch data. Standard errors are reported in parentheses. IDLFM: the proposed individualized dynamic latent factor model, SS: smoothing spline, RNN: recurrent neural network, DRNN: deep recurrent neural network}
\label{tab:real}

\begin{tabular}{c|ccccc}
\hline
\hline
MSE      & IDLFM (Proposed)               & SS            & RNN           & DRNN          \\
\hline
Training & {0.149 (0.011)} & 0.149 (0.013) & 0.701 (0.296) & 0.313 (0.158) \\
Testing  & {0.275 (0.009)} & 0.367 (0.198) &  0.713 (0.144)  & 0.992 (0.666) \\

\hline
\hline
\end{tabular}
\end{center}
\end{table}

\begin{figure}[ht]
\centering
\includegraphics[width=.8\linewidth]{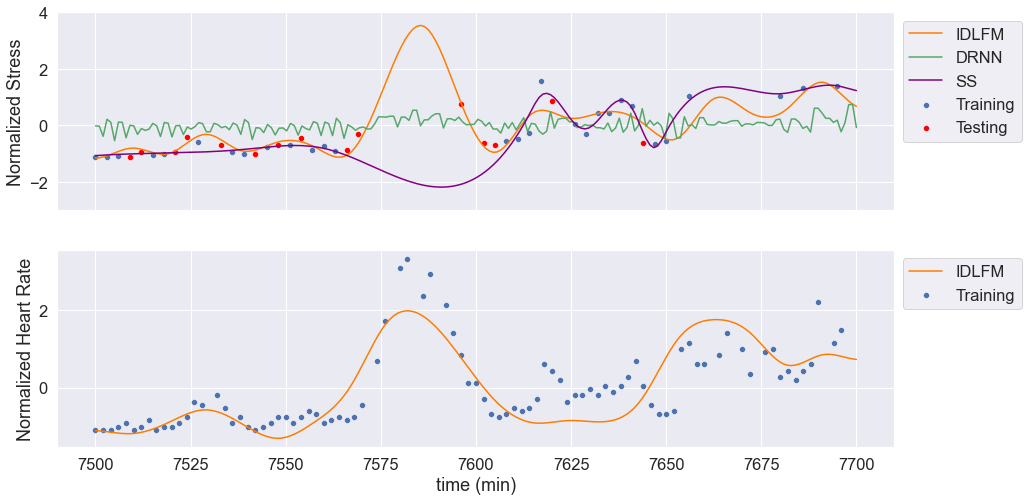}
\caption{The top plot provides estimated normalized stress for one caregiver by the proposed individualized  dynamic latent factor model (orange line) and four competing methods for both training and testing data. The bottom figure provides a fitted normalized heart rate for the same caregiver by the proposed method. Observations of the caregiver in the training set are marked as blue dots, while the ones in the testing set are marked as red dots.}
\label{fig:realdata_fitted}
\end{figure}

Figure \ref{fig:realdata_fitted} illustrates that the proposed individualized  dynamic latent factor model interpolates the stress of caregivers better than competing methods. In particular, estimated normalized stress for one caregiver by the proposed individualized  dynamic latent factor model and competing methods along with the fitted normalized heart rate for the same caregiver by the proposed method are provided using the training and testing data.   The outstanding performance of interpolation power by the proposed method is due to the integration of information across multiple time-series measurements with multi-resolution. For example, the proposed method can identify the high stress in time interval $[7560, 7610]$ based on borrowing information from increased heart rate during the same time period, while the competing methods such as deep recurrent neural network fail to predict a such trend. In practice, monitoring the stress level and identifying high-stress moments is the first step in managing the caregiver's stress level. The proposed method can provide more precise interpolation for the unobserved stress levels, and thus is able to provide effective interventions for caregivers and lessen stress levels during dementia caregiving.

\section{Discussion}
\label{sec:discussion}
We point out two future research directions. The first one is to establish a general algorithm for different approximation methods, since the proposed algorithm are based on the B-spline approximation which may not be applicable for kernel methods. Secondly, in clinical applications, the ultimate goal is to carry out medical tasks such as outcome prediction and patient sub-typing. Interpolating unsampled values and then processing downstream tasks may lead to the suboptimal analyses and predictions \citep{wells2013strategies}. Thus, a potential direction is to extend the proposed interpolation model in prediction, clustering, and classification problems and process the downstream tasks directly based on modelling the time series with missing data.

\section*{Acknowledgement}
The study was supported by 1) Arthur N. Rupe Foundation Caregiver Grant and 2) the University of California, Irvine (UCI) Sue \& Bill Gross School of Nursing Bridge Research Grant. The study was also partially supported by the National Institute of Health (NIH)/National Institute on Aging (NIA) through R01AG069074, and NSF Grant DMS 1952406, DMS 2210640, and DMS 2210860.

This information or content and conclusions are those of the authors and should not be construed as the official position or policy of, nor should any endorsements be inferred by NIH/NIA/NSF or the U.S. Government.

The authors would like to thank the editor, the associate editor and the anonymous referees for providing constructive comments and suggestions that improved the quality of the paper.

\section*{Supplementary Materials}

The supplementary materials provide more simulations, additional assumptions, and proofs of the theorems.

\subsection*{More Simulations}

\subsection{Simulations When $J=101$
}
The three settings of the observation points in training sets when $J=101$ are as follows:

\textbf{Setting 3.1:}
\begin{align}
&Y_{i1}, \cdots, Y_{i,80}: \mathbb{P}(t \in \mathbb{T}_{ij}) = 0.8 \ \text{for} \  t=1,\cdots,T; j=1,\cdots,80, \nonumber \\
&Y_{i,81}, \cdots, Y_{i,101}: \mathbb{P}(t \in \mathbb{T}_{ij}) = 0.2 \ \text{for} \  t=1,\cdots,T;j=81,\cdots,101. \nonumber
\end{align}

\textbf{Setting 3.2:}
\begin{align}
&Y_{i1}, \cdots, Y_{i,60}: \mathbb{T}_{ij} = \{1,\cdots,1000\} \ \text{for} \  j=1,\cdots,60, \nonumber \\
&Y_{i,61}, \cdots, Y_{i,80}: \mathbb{T}_{ij} = \{1,3,5,7,\cdots, 999\}\ \text{for} \  j=61,\cdots,80, \nonumber \\
&Y_{i,81}, \cdots, Y_{i,101}: \mathbb{T}_{ij} = \{1,5,9,13,\cdots, 997\}\ \text{for} \  j=81,\cdots,101. \nonumber 
\end{align}

\textbf{Setting 3.3:}
$$
Y_{i1}, \cdots, Y_{i,101}: \mathbb{P}(t \in \mathbb{T}_{ij}) = 0.7 \ \text{for} \  t=1,\cdots,T; j=1,\cdots,101.
$$

Table \ref{tab:sim3} shows that the proposed method attains lower standard deviation when $J$ is higher due to the integration of multiple time series. The other results are similar to the case when $J=5$.

\begin{table}[ht]
\caption{The mean square error of the proposed method and competing methods under Settings 3.1-3.3 with $J=101$. The standard errors are provided in parentheses. IDLFM: individualized dynamic latent factor model, SS: smoothing spline, RNN: recurrent neural network, DRNN: deep recurrent neural network}
\label{tab:sim3}
\resizebox{\linewidth}{!}{%
\begin{tabular}{c|cc|cc|cc}
\hline
\hline

MSE   & \multicolumn{2}{c|}{Setting 3.1}                   & \multicolumn{2}{c|}{Setting 3.2}                   & \multicolumn{2}{c}{Setting 3.3}                   \\
\hline
      & Training                 & Testing                   & Training                 & Testing                   & Training                 & Testing                   \\
\hline
\begin{tabular}[c]{@{}c@{}}IDLFM\\ (Proposed)\end{tabular} & \textbf{0.253 (0.026)} & \textbf{0.260 (0.027)} & \textbf{0.251 (0.005)} & \textbf{0.257 (0.007)} & \textbf{0.250 (0.003)} & \textbf{0.255 (0.006)} \\
SS                                                         & 0.561 (0.276)          & 2.842 (9.136)          & 0.569 (0.257)          & 0.616 (0.342)          & 0.541 (0.260)          & 0.585 (0.355)          \\
RNN                                                        & 8.797 (7.788)          & 8.787 (7.768)          & 8.081 (8.710)          & 8.218 (8.852)          & 6.356 (5.947)          & 6.346 (5.941)          \\
DRNN                                                       & 4.532 (3.761)          & 4.528 (3.740)          & 2.291 (2.569)          & 5.795 (7.375)          & 1.345 (1.270)          & 1.347 (1.270)                  \\

\hline
\hline
\end{tabular}%
}
\end{table}

\subsection{Simulations Under Various Missing Mechanisms}

To account for the robustness of the proposed method under various missing mechanisms, we conducted the following additional simulations. The data generation procedures in all three settings are the same as Section 5.1. Three missing mechanisms are considered: Missing Completely at Random (MCAR), Missing at Random (MAR), and Missing Not at Random (MNAR). For Missing Completely at Random (the same as Setting 1.3 in the revised manuscript), missingness occurs randomly and is not associated with any other variables. In the setting Missing at Random (MAR), the missingness of $Y_J$ is dependent on the value of $Y_1$, which is fully observed. Specifically, observations of $Y_J$ are missing when $Y_1$ exceeds the $90\%$ quantile of $Y_1$. The missing pattern is driven by observed variables, making it "missing at random" in relation to $Y_1$. As for Missing Not at Random (MNAR), missingness depends on the unobserved values, i.e., observations of $Y_J$ are missing when $Y_J$ itself exceeds the $90\%$ quantile of $Y_J$.

\begin{table}[h!]
\begin{center}
\caption{The mean squared error of the proposed method under different missing mechanisms. The standard errors are provided in parentheses. MCAR: missing completely at random. MAR: missing at random. MNAR: missing not at random.}
\label{tab:missingness}

\begin{tabular}{c|cc}
\hline
\hline
               & Training                 & Testing                   \\
\hline
MCAR           & 0.257 (0.028) &    0.261 (0.025) \\
MAR            & 0.262 (0.040)          & 0.262 (0.032)          \\
MNAR           & 0.236 (0.033)          & 0.287 (0.046)          \\

\hline
\hline
\end{tabular}
\end{center}
\end{table}
The results presented in Table \ref{tab:missingness} illustrate the robustness of the proposed method under three missing mechanisms. The interpolation error under missing at random is almost the same as the one under missing completely at random. In the case of missing not at random, the mean squared error is only marginally higher, by around $10\%$, when compared to the scenario of missing completely at random.

\subsection{Simulations of Selecting Rank $R$}

The selected rank through cross-validation from 100 simulations under Settings 1.1-1.3 in Section 5 are presented in Table \ref{tb:cross}. The true rank of 3 is most frequently selected among the rank candidates. This finding also suggests that the cross-validation approach is effective and reliable for the proposed method.

\begin{table}[h!]
\begin{center}
\caption{The count of rank chosen by cross-validation in 100 simulations, where the true rank is $3$. }
\label{tb:cross}
\begin{tabular}{c|ccccc}
\hline
\hline
                   R    & Setting 1.1    & Setting 1.2    & Setting 1.3 \\
\hline
                1       &  1             & 0            & 0            \\
                2       & 10             & 3            & 4                \\
            \textbf{3}  & \textbf{56}    & \textbf{48}  & \textbf{40}       \\
                4       & 24             & 27           & 28                       \\
                5       & 5          & 11          & 17                       \\
                  6   & 4          & 11          & 11   \\
\hline
\hline
\end{tabular}
\end{center}
\end{table}

To gain more insight on the performance of the proposed method under different ranks, we conducted an additional simulation. In this simulation, we adopt the same settings as in Setting 1.1-1.3 in Section 5, where the true rank is 3. On the test sets, we compare the performance of the proposed method among ranks chosen from 1 to 6.

\begin{table}[h!]
\begin{center}
\caption{The mean squared error on the test set of the proposed method under different ranks $R$ where the true rank is $3$. The standard errors are provided in parentheses. }
\label{tab:R}
\begin{tabular}{c|ccccc}
\hline
\hline
                   R   & Setting 1.1       & Setting 1.2        & Setting 1.3 \\
\hline
 1 & 1.254 (0.689)          & 1.136 (0.693)          & 0.978 (0.685)                       \\
                 2 & 0.493 (0.182)          & 0.475 (0.167) & 0.430 (0.170)                       \\
                    \textbf{3} & \textbf{0.415 (0.120)} & \textbf{0.376 (0.080)}          & \textbf{0.344 (0.050)}              \\
                    4 & 0.449 (0.129)          & 0.383 (0.076)          & 0.345 (0.043)                       \\
                    5 & 0.458 (0.138)          & 0.404 (0.060)          & 0.353 (0.043)                       \\
                    6 & 0.470 (0.129)          & 0.405 (0.061)          & 0.355 (0.042)   \\
\hline
\hline
\end{tabular}
\end{center}
\end{table}

According to simulation results in Table \ref{tab:R}, the proposed method performs the best using the true rank under all three settings.  Moreover, the mean squared errors on the test sets are close to those associated with the true rank, when the rank is larger than the true rank. This might be the reason that alternative selections in Table \ref{tb:cross} tend to lean towards a higher rank of 4, while a rank smaller than 3 is chosen less infrequently. 

\subsection{Simulation of Downstream Prediction Analysis}

We demonstrate the practical benefits of the proposed method for downstream analysis through the following simulation. This simulation specifically highlights how the proposed method can improve prediction in regression tasks.

In this simulation, four time series are generated as X, and one time series (Y) was generated as the target variable. The data is generated under Setting 1.1 in Section 5, where the samples are collected under the random multi-resolution case. The downstream task is to predict $Y$ based on $X$ via a linear regression model. We compare the performance of the regression model before and after applying the proposed interpolation method based on the mean squared error, defined on a set $\Omega$ as $\sum_{k \in \Omega} \{Y(t_k)-\hat Y(t_k)\}^2/|\Omega|$. First, we fit a regression model to predict Y using X with the original data, where only time points with complete observations were used. Second, we fit the regression model using the interpolated data and assess its prediction error in a similar manner. Additionally, for reference, the regression MSE based on the full data (without any missing values) is also calculated.

\begin{table}[h!]
\begin{center}
\caption{The prediction error (based on MSE) of linear regression before interpolation, after interpolation, and based on complete data. The standard errors are provided in parentheses. }
\begin{tabular}{c|c|ccc}
\hline
\hline
  &     & Before Interpolation & After Interpolation  & Full Data  \\
\hline
\multirow{3}{*}{T} & 50  & 11.960 (57.151)        & \textbf{0.516 (0.267)} &0.433 (0.219) \\
  & 100 & 0.847 (0.595)        & \textbf{0.573 (0.270)}  &0.512 (0.239)\\
  & 200 & 0.627 (0.272)        & \textbf{0.524 (0.185)} &0.482 (0.172) \\
\hline
\hline
\end{tabular}
\end{center}
\end{table}

Our numerical study provides a significant reduction of at least $20\%$ in prediction error following the interpolation performed by our proposed method. This improvement is particularly pronounced when the sample size is limited. Furthermore, in cases where the sample sizes are large, the prediction error after interpolation approaches closely to that obtained with full data.  For instance, when $T = 200$, the mean squared error using interpolated data is approximately $8\%$ worse than the MSE derived from the full data, compared to a difference of $19\%$ when $T = 50$.

\subsection{Simulations on Extrapolation of Proposed Method}

The proposed method is designed for interpolation, but it can potentially be applied to extrapolation, albeit for a limited duration. To demonstrate the extrapolation ability, we conduct additional simulations, mirroring Setting 1.1. In this setting, instead of generating $T$ time points, we extend it to $T + 10$ points.  After training the model with data from 0 to $T$, we predict  $Y_{iJ}$ at $T+1$, $T+3$, $T+6$, and $T+10$ by 

$$
    \widehat{Y}_{ij}(t) = \widehat{\boldsymbol{f}}_j^T \widehat{\boldsymbol{W}}_i \boldsymbol{B}(t),
 \; \text{for} \; t = T+1 ,T+3,T+6,  \; \text{or} \; T+10.
$$
We compare the performance of the proposed method for interpolation and extrapolation tasks based on the mean squared error (MSE), defined at time $t$ as $\sum_{ij} \{Y_{ij}(t)-\hat Y_{ij}(t)\}^2/(IJ)$.

\begin{table}[h!]
\begin{center}
\caption{The interpolation extrapolation errors of the proposed method based on mean squared error (MSE) of the proposed method. The standard errors are provided in parentheses. }
\label{tab:prediction}
\begin{tabular}{c|c|c|c|c|c}
\hline
\hline
    & $\le T$ & $T+1$     & $T+3$ & $T+6$     & $T+10$           \\
    \hline
MSE & 0.488 (0.188)  & 0.682 (0.353) & 3.966 (4.323) & 10.966 (13.813) & 13.700 (18.433) \\
\hline
\hline
\end{tabular}
\end{center}
\end{table}

As shown in Table \ref{tab:prediction}, the prediction error observed at $T+1$ closely resembles the interpolation error, whereas the prediction error at $T+10$ is notably larger. This discrepancy indicates that the proposed approach is applicable for extrapolation, but reliable only over a limited time span. The constraints on the extrapolation capability are due to employing the B-spline approximation method directly for extrapolation purposes. To improve the performance in extrapolation, one might consider alternative methods which are more suited for extrapolation tasks, such as the partitioning methods explored in \citet{strelkovskaya2020spline}. However, this direction is not the primary focus of our paper.

\subsection{Real Data Analysis Adjusted by Baseline Covariate}

In the caregiver study, baseline covariates concerning caregivers are available, which include gender, age, relationship to the dementia patient, primary language spoken at home, and household income. These covariates can be used in the estimation, but its improvement of interpolation error is negligible based on the following analysis of the real data.

 To adjust baseline covariates, we conduct simple linear regressions with stress level and the other four variables as responses, and gender, age, relationship to the dementia patient and spoken language as baseline variables. We choose these baseline variables according to their potential effects on stress. The proposed method is subsequently employed on the residuals obtained from this regression. The mean squared errors of estimated standardized stress by the proposed method are computed as in Section 6, when stress is adjusted and not adjusted by baseline variables, respectively. 

\begin{table}[h!]
\begin{center}
\caption{The mean squared errors of the proposed estimator based on adjusted or unadjusted stress by demographic variables (gender, age, relationship to dementia patient, language). The standard errors are provided in the parentheses.}
\label{tab:covariate}

\begin{tabular}{c|cc}
\hline
\hline
               & Training                 & Testing                   \\
\hline
Adjusted           & 0.1983 (0.0025) &    0.2667 (0.0032) \\
Not Adjusted            & 0.1984 (0.0024)          & 0.2667 (0.0032)          \\
\hline
\hline
\end{tabular}
\end{center}
\end{table}

Table \ref{tab:covariate} shows the mean squared errors of estimated stress based on the proposed method are almost the same, whether the additional covariates are incorporated or not. This implies that the baseline covariates might not significantly contribute to the interpolation of observed samples. 

\subsection*{Additional Assumptions for Theorem \ref{thm:convgen} and Theorem \ref{thm:convB}}

We  introduce the additional assumptions for  Theorem \ref{thm:convgen} and Theorem \ref{thm:convB} in this section.

\begin{assumption}
\label{asm:epsilon}
For random noises $\epsilon_{ij}(t)$ $(i = 1,\cdots, I, j = 1, \cdots, J, t \in [0,T])$, we assume
$$
E[exp\{c_1 \epsilon_{ij}(t)\}] \leq \infty,
$$
for a constant $c_1$.

Assumption \ref{asm:epsilon} yields the exponential probability bound. But such assumption is not necessary for the rate of convergence.

\end{assumption}
\begin{assumption}
\label{asm:designmatrix}
For fixed observation time points $\mathbb{T}_{ij} $, we assume the smallest positive eigenvalue of matrix $K_{ij}^{-1}H_{ij}^TH_{ij}$ is bounded below by a constant
$g_0 > 0$, where
$$
H_{ij} = \left( \begin{matrix}
1 & \cdots & t_{ij1} & \cdots & t_{ij1}^\alpha \\
1 & \cdots & \cdots  & \cdots & \cdots \\
1 & \cdots & \cdots & \cdots & \cdots \\
1 & \cdots & t_{ijK_{ij}} & \cdots & t_{ijK_{ij}}^\alpha
\end{matrix} \right). 
$$
\end{assumption}

\subsection*{Proof of Theorem \ref{thm:convgen}}

Let  $l_{\Delta}(\boldsymbol{\Psi} \mid \cdot)=l(\boldsymbol{\Psi}, \cdot)-l\left(\boldsymbol{\Psi}_{0}, \cdot\right)$, and
$$
K\left(\boldsymbol{\Psi}, \boldsymbol{\Psi}_{0}\right)=\frac{1}{\sum K_{ij}} \sum_{i=1}^{I} \sum_{j = 1}^{J} \sum_{k = 1}^{K_{ij}} E\left[l_{\Delta}\left\{\boldsymbol{\Psi}, Y_{ij}(t_k)\right\}\right],
$$
which is the expected loss difference between $\boldsymbol{\Psi}$ and $\boldsymbol{\Psi}_{0}$. Since $\boldsymbol{\Psi}_{0}$ is the true parameter, we have $K\left(\boldsymbol{\Psi}_{0},\boldsymbol{\Psi}\right) \geq 0$ for all $\boldsymbol{\Psi} \in \mathcal{S}$ and $K=0$ only if $\boldsymbol{\Psi}=\boldsymbol{\Psi}_0$. Given $K\left(\boldsymbol{\Psi}, \boldsymbol{\Psi}_{0}\right)$, the distance between $\boldsymbol{\Psi}$ and $\boldsymbol{\Psi}_0$ can be measured as $\rho\left(\boldsymbol{\Psi}_0,\boldsymbol{\Psi}\right)=K^{1 / 2}\left(\boldsymbol{\Psi}_0,\boldsymbol{\Psi}\right)$. Similarly, we quantify the variance of the loss difference $l_{\Delta}$ as:
\begin{equation}
\begin{aligned}
  &V\left(\boldsymbol{\Psi}_0,\boldsymbol{\Psi}\right)=\\
  &\frac{1}{\sum K_{ij}} \sum_{i=1}^{I} \sum_{j = 1}^{J} \sum_{k = 1}^{K_{ij}} \operatorname{Var}\left[l_{\Delta}\left\{\boldsymbol{\Psi}, Y_{ij}(t_k)\right\}\right]. \nonumber
\end{aligned}
\end{equation}

We first prove a lemma related to the error bound for observed time points, then extend the result to the average integrated error in Theorem \ref{thm:convgen}.
\begin{lemma}
\label{lm:errorgen}
Under Assumptions \ref{asm:epsilon}, \ref{asm:designmatrix}, we have 
\begin{equation}
\label{eq:Qnresult}    P\left(L(\boldsymbol{\Psi}_0| \boldsymbol{Y}) - \inf_{\rho(\boldsymbol{\Psi}, \boldsymbol{\Psi}_0) > \eta_{|\Omega|}, \boldsymbol{\Psi} \in \mathcal{S}} L(\boldsymbol{\Psi}| \boldsymbol{Y}) > -\eta_{|\Omega|}^2/2\right)  \leq 7 \exp \left(-c_{2}|\Omega| \eta_{|\Omega|}^{2}\right),
\end{equation}
where $c_{2} \geq 0$ is a constant, $\eta_{|\Omega|} \sim |\Omega|^{-\alpha/(2\alpha+1) }$.
\end{lemma}
\color{black}
\begin{proof}
For any $k_i \geq 0$, let $A\left(k_{1}, k_{2}\right)=\left\{\boldsymbol{\Psi} \in \mathcal{S}: k_{1} \leq \rho\left(\boldsymbol{\Psi}_0,\boldsymbol{\Psi}\right) \leq 2 k_{1}, J(\boldsymbol{\Psi}) \leq k_{2}\right\}$, and $\mathcal{F}\left(k_{1}, k_{2}\right)=\left\{l_{\Delta}(\boldsymbol{\Psi} \mid \cdot): \boldsymbol{\Psi} \in A\left(k_{1}, k_{2}\right)\right\}$.

We verify several conditions of Theorem 2 in \citep{shen1998method}. First, we verify Assumption B. By definition, we have $\operatorname{E}\left\{l_{\Delta}\left(\boldsymbol{\Psi}, Y_{ij}(t)\right)\right\} = (\psi_{ij}-\psi_{0,ij})^2$, and $\operatorname{Var}\left\{l_{\Delta}\left(\boldsymbol{\Psi}, Y_{ij}(t)\right)\right\} = 4(\psi_{ij}-\psi_{0,ij})^2\sigma^2$. Thus $\sup _{A\left(k_{1}, k_{2}\right)} V\left(\boldsymbol{\Psi}_0,\boldsymbol{\Psi}\right) \leq c_{3} k_{1}^{2}=c_{3} k_{1}^{2}\left\{1+\left(k_{1}^{2}+k_{2}\right)^{\beta_{1}}\right\},$ and hence $\beta_{1}=0$. In the rest of this section, all $c_{i}$ 's with $i \in \mathbb{N}$ are assumed to be non-negative constants. 

Second, for Assumption  C,  $\left|l\left(\boldsymbol{\Psi}, Y_{ij}\right)-l\left(\boldsymbol{\Psi}_0, Y_{ij}\right)\right|=\left|\psi_{ij}-\psi_{0,ij}\right| \cdot\left|2 Y_{ij}-\psi_{0, ij}-\psi_{ij}\right|$. Define a new random variable $w=\left|2 Y_{ij}-\psi_{0, ij}-\psi_{ij}\right|,$ then we have $\mathrm{E}\left\{\exp \left(t_{0} w\right)\right\}<\infty$ for $t_{0}$ at an open interval containing 0 by Assumption \ref{asm:epsilon}.

Now we verify that for a constant $c_4>0,$ we have $\sup _{A\left(k_{1}, k_{2}\right)} \|\boldsymbol{\Psi}_{0}-\boldsymbol{\Psi} \|_{\text {sup }} \leq c_{4}\left(k_{1}^{2}+k_{2}\right)^{\beta_{2}}$ for $\beta_{2} \in[0,1) .$ Since $\psi_{ij}(t) \in W_{q}^{\alpha}[0, T]$, by Lemma 2 in \citet{shen1998method}, $\sup _{A\left(k_{1}, k_{2}\right)} \|\psi_{ij}-\psi_{0,ij}\|_{\text {sup }} \leq k_1^a k_2^{1-a}$, where $a = (\alpha - 1/q)/(\alpha - 1/q+1/2)$. Since $k_1^a k_2^{1-a} \leq c_4 \left(k_{1}^{2}+k_{2}\right)^{1-a/2}$, we have $\beta_2 = 1-a/2$.

Next, we verify the Assumption D. Let
$$
\mathcal{N}(\varepsilon, n)=\left\{g_{1}^{l}, g_{1}^{u}, \ldots, g_{n}^{l}, g_{n}^{u}\right\},
$$
be a set of functions from the $L_{2}$ space, where $\max _{1 \leq i \leq n}\left\|g_{i}^{u}-g_{i}^{l}\right\|_{2} \leq \varepsilon$.
Suppose for any function $l_{\Delta} \in \mathcal{F}\left(k_{1}, k_{2}\right),$ there exists $i \in\{1, \ldots, n\}$ such that $g_{i}^{l} \leq l_{\Delta} \leq g_{i}^{u}$ almost surely. Then the Hellinger metric entropy is defined as $H(\varepsilon, \mathcal{F})=\log \{n: \min \mathcal{N}(\varepsilon, n)\}$.
 Define
$$
\psi\left(k_{1}, k_{2}\right)=\int_{L_{0}}^{U_{0}} H^{1 / 2}(u, \mathcal{F}) d u / L_{0},
$$
where $L_{0}=c_{5} \lambda_{|\Omega|}\left(k_{1}^{2}+k_{2}\right)$ and $U_{0}=c_{6} \varepsilon_{|\Omega|}\left(k_{1}^{2}+k_{2}\right)^{\left(1+\max \left(\beta_{1}, \beta_{2}\right)\right) / 2} .$ Based
on Theorem 5.2 of \citep{birman1967piecewise}. the Hellinger metric entropy is controlled by
$$
H\left(\varepsilon_{|\Omega|}, \mathcal{F}\right) \leq c_{7} \varepsilon_{|\Omega|}^{-1/\alpha}.
$$
Note that the Assumption \ref{asm:designmatrix} is needed here to make sure that $c_{7} < \infty$, refer to \citet{shen1998method} for the detailed arguments.
Then for fixed $k_{1}$ and $k_{2},$ we have $\psi\left(k_{1}, k_{2}\right)\leq$ $c_{8} \frac{U_{0}^{1-1/2\alpha}}{L_{0}} \sim \frac{\varepsilon_{|\Omega|}^{1-1/2\alpha}}{\lambda_{|\Omega|}} .$ Given that $\psi \left(k_{1}, k_{2}\right)\sim|\Omega|^{1 / 2},$ the best possible rate is
achieved at $\varepsilon_{|\Omega|} \sim \lambda_{|\Omega|}^{1 / 2},$ that is,
$$
\varepsilon_{|\Omega|} \sim |\Omega|^{-\frac{\alpha} {2\alpha+1} }.
$$
 By applying Theorem 2 of \citet{shen1998method}, we have the result in Lemma \ref{lm:errorgen} when $\eta_{|\Omega|}=\max \left(\varepsilon_{|\Omega|}, \lambda_{|\Omega|}^{1 / 2}\right)$, and $\varepsilon_{|\Omega|} \sim |\Omega|^{-\alpha/(2\alpha+1) }$ is the best possible rate that can be achieved when $\lambda_{|\Omega|} \sim \varepsilon_{|\Omega|}^{2}$.
\end{proof}

We then utilize Lemma \ref{lm:errorgen} to prove Theorem \ref{thm:convgen}. Let $s_{ij}(\cdot) =  \left\{ \psi_{ij}(\cdot)-\psi_{0,ij}(\cdot)\right\}^2$. Since $K_{ij} \sim K$, we have,
\begin{equation}
\label{eq:phistar}
\begin{aligned}
   & \sqrt{\frac{1}{IJ} \sum_{i=1}^{I} \sum_{j = 1}^{J} \int_{0}^{T}s_{ij}(t)dQ_{ij}(t)} \\
    \le & \sqrt{\frac{1}{IJ} \sum_{i=1}^{I} \sum_{j = 1}^{J} \int_{0}^{T}s_{ij}(t)dQ_{ij,n}(t)}+  \sqrt{\frac{1}{IJ} \sum_{i=1}^{I} \sum_{j = 1}^{J} \left| \int_{0}^{T}s_{ij}(t)d(Q_{ij}-Q_{ij,n})(t) \right|}\\
    = & \sqrt{\frac{1}{IJ} \sum_{i=1}^{I} \sum_{j = 1}^{J} \frac{1}{K_{ij}} \sum_{k=1}^{K_{ij}}s_{ij}(t_k)}+  \sqrt{\frac{1}{IJ} \sum_{i=1}^{I} \sum_{j = 1}^{J} \left| \int_{0}^{T}s_{ij}(t)d(Q_{ij}-Q_{ij,n})(t) \right|}\\
    \le & c_9\sqrt{\frac{1}{\sum K_{ij}} \sum_{i=1}^{I} \sum_{j = 1}^{J} \sum_{k = 1}^{K_{ij}} s_{ij}(t_k)}  + \sqrt{\frac{1}{IJ} \sum_{i=1}^{I} \sum_{j = 1}^{J} \left| \int_{0}^{T}s_{ij}(t)d(Q_{ij}-Q_{ij,n})(t) \right|}.
\end{aligned}
\end{equation}
for some constant $c_9$.

Note that the first term in the last line is $c_{9}\rho(\boldsymbol{\Psi}, \boldsymbol{\Psi}_0)$. However, the above result cannot directly be applied to $\rho(\widehat{\boldsymbol{\Psi}}, \boldsymbol{\Psi}_0)$, as $E(l_{\Delta}(\widehat{\boldsymbol{\Psi}}|Y_{ij}(t_k))) = \left\{ \hat \psi_{ij}(\cdot)-\psi_{0,ij}(\cdot)\right\}^2$ is not necessarily true. Thus, we need some tricks to extend the result to $\rho(\widehat{\boldsymbol{\Psi}}, \boldsymbol{\Psi}_0)$.

Define 
$$
\begin{aligned}
    &\rho^{\star}(\boldsymbol{\Psi}, \boldsymbol{\Psi}_0) \\
    =& \sqrt{\frac{1}{IJ} \sum_{i=1}^{I} \sum_{j = 1}^{J} \int_{0}^{T}\left\{ \psi_{ij}(\cdot)-\psi_{0,ij}(\cdot)\right\}^2dQ_{ij}(t)} \\
    &- \sqrt{\frac{1}{IJ} \sum_{i=1}^{I} \sum_{j = 1}^{J} \left| \int_{0}^{T}\left\{ \psi_{ij}(\cdot)-\psi_{0,ij}(\cdot)\right\}^2d(Q_{ij}-Q_{ij,n})(t) \right|}.
\end{aligned}
$$
Note that $\rho(\boldsymbol{\Psi}, \boldsymbol{\Psi}_0) = \sqrt{\frac{1}{\sum K_{ij}} \sum_{i=1}^{I} \sum_{j = 1}^{J} \sum_{k = 1}^{K_{ij}} s_{ij}(t_k)}$. Then, by (\ref{eq:phistar}), we have $c_9\rho(\boldsymbol{\Psi}, \boldsymbol{\Psi}_0) \ge \rho^{\star}(\boldsymbol{\Psi}, \boldsymbol{\Psi}_0)$.

Additionally, by (\ref{eq:assumptheta}), if $\rho^{\star}(\widehat{\boldsymbol{\Psi}}, \boldsymbol{\Psi}_0) \ge \eta_{|\Omega|}$, we have \begin{equation}
\label{eq:1}
\begin{aligned}
        &\inf_{\rho^{\star}(\boldsymbol{\Psi}, \boldsymbol{\Psi}_0) \ge \eta_{|\Omega|}, \boldsymbol{\Psi} \in \mathcal{S}}  L(\boldsymbol{\Psi}| \boldsymbol{Y}) \\
        \le &L(\widehat{\boldsymbol{\Psi}}| \boldsymbol{Y}) \\
        \le &\inf_{ \boldsymbol{\Psi} \in \mathcal{S}}  L(\boldsymbol{\Psi}| \boldsymbol{Y}) + \tau_{|\Omega|} \\
        \le &L(\boldsymbol{\Psi}_0| \boldsymbol{Y}) + \tau_{|\Omega|}.
\end{aligned}
\end{equation}

Thus,
$$
\begin{aligned}
    &P(\rho^{\star}(\widehat{\boldsymbol{\Psi}}, \boldsymbol{\Psi}_0) \ge \eta_{|\Omega|}) \\
    \le &P\left(L(\boldsymbol{\Psi}_0| \boldsymbol{Y}) -\inf_{\rho^{\star}(\boldsymbol{\Psi}, \boldsymbol{\Psi}_0) \ge \eta_{|\Omega|}, \boldsymbol{\Psi} \in \mathcal{S}}  L(\boldsymbol{\Psi}| \boldsymbol{Y}) \ge -\tau_{|\Omega|}\right) \\
    \le &P\left(L(\boldsymbol{\Psi}_0| \boldsymbol{Y}) -\inf_{\rho^{\star}(\boldsymbol{\Psi}, \boldsymbol{\Psi}_0) \ge \eta_{|\Omega|}, \boldsymbol{\Psi} \in \mathcal{S}}  L(\boldsymbol{\Psi}| \boldsymbol{Y}) \ge -\eta_{|\Omega|}^2/(2c_9^2)\right) \\
    \le &P\left(L(\boldsymbol{\Psi}_0| \boldsymbol{Y}) -\inf_{c_9\rho(\boldsymbol{\Psi}, \boldsymbol{\Psi}_0) \ge \eta_{|\Omega|}, \boldsymbol{\Psi} \in \mathcal{S}}  L(\boldsymbol{\Psi}| \boldsymbol{Y}) \ge -\eta_{|\Omega|}^2/(2c_9^2)\right).
\end{aligned}
$$

Thus, $\rho^{\star}(\widehat{\boldsymbol{\Psi}}, \boldsymbol{\Psi}_0) = O_p(\eta_{|\Omega|})$ by Lemma \ref{lm:errorgen}.

Note that by integration by parts and the fact that $(Q_{ij,n}-Q_{ij})(T) = (Q_{ij,n}-Q_{ij})(0) = 0$, and let $h_{ij}(\cdot) =  \left\{ \hat{\psi}_{ij}(\cdot)-\psi_{0,ij}(\cdot)\right\}^2$, we have
$$
\begin{aligned}
    &\left| \int_{0}^{T}h_{ij}(t)d(Q_{ij,n}-Q_{ij})(t) \right| \\
    = & \left| h_{ij}(t)(Q_{ij,n}-Q_{ij})(t) \bigg|_{0}^{T}-\int_{0}^{T}(Q_{ij,n}-Q_{ij})(t)h^{(1)}_{ij}(t)dt \right| \\
    \le &  o(K^{-1})\int_{0}^{T} |h^{(1)}_{ij}(t)|dt,
\end{aligned}
$$
where $f^{(1)}(\cdot)$ is the first derivative of $f(\cdot)$.

Note that 
$$
\begin{aligned}
    &\int_{0}^{T} |h^{(1)}_{ij}(t)|dt \\
    =&\int_{0}^{T} 2 \left|\hat \psi_{ij}(t)-\psi_{0,ij}(t)\right| \left|\hat \psi^{(1)}_{ij}(t)-\psi^{(1)}_{0,ij}(t)\right|dt \\
    \le &2 \|\hat \psi_{ij}(t)-\psi_{0,ij}(t)\|_{L_2} \|\hat \psi^{(1)}_{ij}(t)-\psi^{(1)}_{0,ij}(t)\|_{L_2}.
\end{aligned}
$$
So $\int_{0}^{T} |h^{(1)}_{ij}(t)|dt$ is bounded since $\hat \psi_{ij}(t), \psi_{0.ij}(t) \in W_{q}^{\alpha}[0, T]$.

Thus, 
$$
\sqrt{\frac{1}{IJ} \sum_{i=1}^{I} \sum_{j = 1}^{J} \int_{0}^{T}h_{ij}(t)dQ_{ij}(t)} = O_{p}\{(IJK)^{-\alpha/(2\alpha+1) }\} + o_{p}(K^{-1/2}),
$$
since $|\Omega| \sim IJK$ by $K_{ij} \sim K$.

\subsection*{Proof of Theorem \ref{thm:convB}}
\begin{proof}
Note that $\mathcal{S}_M \subset \mathcal{S}$, so the argument in Theorem \ref{thm:convgen} is valid when replacing $\psi_{ij}$ to $\phi_{ij}$. Additionally, the Theorem 2 of \citet{shen1998method} also applicable since the result is rely on Theorem 3 of \citet{shen1994convergence}, which utilize the sample estimator satisfying \eqref{eq:assumpthetanew}. Then by Lemma \ref{lm:errorgen}, we have
\begin{equation}
\label{eq:Qnresultnew}    P\left(L(\boldsymbol{\Psi}_0| \boldsymbol{Y}) - \inf_{\rho(\boldsymbol{\Phi}, \boldsymbol{\Psi}_0) > \eta_{|\Omega|}, \boldsymbol{\Phi} \in \mathcal{S}_M} L(\boldsymbol{\Phi}| \boldsymbol{Y}) > -\eta_{|\Omega|}^2/2\right)  \leq 7 \exp \left(-c_{10}|\Omega| \eta_{|\Omega|}^{2}\right),
\end{equation}
where $c_{10} \geq 0$ is a constant, $\eta_{|\Omega|} \sim |\Omega|^{-\alpha/(2\alpha+1) }$, and  $\lambda_{|\Omega|} \sim \varepsilon_{|\Omega|}^{2}$.

The results in Theorem \ref{thm:convB} then follows by the same argument in Theorem \ref{thm:convgen}. The only thing need to mention is that the last inequation in \eqref{eq:1} holds because when $M$ is large enough, there exists a $\boldsymbol{\Phi}_{0} \in \mathcal{S}_M$ such that $\phi_{0,ij}(t) = \psi_{0,ij}(t)$ for $i,j,t\in \Omega$. Thus, $L(\boldsymbol{\Phi}_0| \boldsymbol{Y}) = L(\boldsymbol{\Psi}_0| \boldsymbol{Y})$.

\end{proof}
\bibliographystyle{apa}
\bibliography{reflist}
\end{CJK*}
\end{document}